\documentclass{article}
\usepackage{cite}
\usepackage{geometry}
\title{ Quantum Field Theory Applications of Heun Type Functions}
\author{ T. Birkandan \\ Istanbul Technical University, Department of Physics, 34469 Istanbul, Turkey \\ e-mail: birkandant@itu.edu.tr  \\[2ex]
         M. Horta\c{c}su \\ Mimar Sinan Fine Arts University, Department of Physics, 34380 Istanbul, Turkey \\ e-mail: hortacsu@itu.edu.tr  }
\begin{document}
\maketitle
\begin{abstract}
After a brief introduction to Heun type functions we note that the actual solutions of the eigenvalue equation emerging in the calculation of the one loop contribution to QCD from the Belavin-Polyakov-Schwarz-Tyupkin instanton and the similar calculation for a Dirac particle coupled to a scalar $CP^1$ model in two dimensions can be given in terms of confluent Heun equation in their original forms. These equations were previously modified to be solved by more elementary functions. We also show that polynomial solutions with discrete eigenvalues are impossible to find in the unmodified equations.
\end{abstract}
\section{Introduction}
Heun function was introduced in 1889 by Karl Heun \cite{Heun}. This function is the solution of differential equations with four regular singularities and any second order differential equation with four regular singularities can be reduced to this form.

Although Heun equation was introduced in 1889, most theoretical physicists were not acquainted with this equation until 1990s. An important point in the history of this equation and its confluent forms is the ``Centennial Workshop on Heun Equations: Theory and Applications, Sept. 3-8, 1989, Schloss Ringberg". The presentations at this workshop were printed in a book edited by A. Ronveaux \cite{Ronveaux3}. After this meeting, this equation became more popular among theoretical physicists. After 1990s, it is encountered in increasingly many papers in theoretical physics. Work done before the year 2000 can be found in the book by Slavyanov and Lay \cite {Slavyanov}. For later applications, especially in general relativity, Horta\c{c}su gives a long list\cite{Hortacsu}.

Note that Heun was not cited in the celebrated work of Teukolsky \cite {Teukolsky} when he wrote  his ``\emph{Teukolsky Master Equations}". These equations were shown to be reducible to a form of the Heun class later. Batic and Schmid \cite{Batic} give references to people who tried to show whether "Teukolsky Master Equations" were one of the confluent forms of the Heun equation \cite{Poons,Leaver,Suzuki}. Then Batic et al., in reference given above, showed that "Teukolsky Master Equation" could be transformed in any physically relevant D type metric into a Heun form. In this paper we will try to identify one equation which was studied by Prof. 't Hooft in 1976, and show that in its original form, it is a confluent form of the Heun equation.

The Heun type solution is not specific for D type metrics. Scalar and spinor wave equations, when written in the background of a Petrov I type instanton metric with two commuting Killing vectors and a Killing tensor also have Heun type solutions \cite{Nutku,Tolga,Tolga2}.

The eigenvalue equation emerging in the calculation of the one loop contribution to QCD \cite{Hooft} from the Belavin-Polyakov-Schwarz-Tyupkin (BPST) instanton \cite{Belavin} and the calculation for a Dirac particle coupled to a scalar $CP^1$ model in two dimensions \cite{Hortacsu1} were shown to yield hypergeometric functions, after these equations were modified slightly. Actually, the original equations give confluent Heun type solutions. These authors do not mention Heun type solutions in their papers.

We should note that it was the remarkable foresight of Gerard 't Hooft to devise a way to obtain a meaningful result from a complicated equation. If the equation was not modified, his result, which was of great importance in those days, would not have been obtained. Here, we will just show that the two unaltered equations are of confluent Heun type.

Polynomial solutions of confluent Heun equation can be obtained if the parameters of the equation satisfy certain conditions \cite {Arscott,Fiziev,hakem,Ciftci,Karayer}. We will show that our quantum field theory examples do not satisfy these conditions and hence they do not have polynomial solutions.
\section{Heun Equation}
Heun equation is the most general second order differential equation with four regular singular points. Usually these singular points are taken to be at zero, one, an arbitrary point $d$ (other
than zero and one) and at infinity. The general equation reads
\begin{equation}
\frac {d^{2}H_g}{dz^{2}} + \bigg[\frac {\gamma}{z} + \frac{ \delta}{z-1}
+ \frac{ \epsilon}{z-d} \bigg] \frac{dH_g}{dz} - \frac{\alpha \beta
z-q}{z(z-1)(z-d)}H_g = 0,
\end{equation}
where $H_g$ is the solution of the Heun's general equation. We have the Fuchs relation between the constants given as $\alpha+\beta+1 = \gamma + \delta + \epsilon$ which ensures the regularity of the singularity at infinity. Here, $q$ is the accessory parameter. The distinctive property of Heun function is not having a two way recursion relation between the coefficients when one expands the solution as an infinite power series around one of its regular singular points. The recursion relation may contain at least three, sometimes more coefficients, making it hard to extract the asymptotic properties from this expansion. Also, \emph{no example has been  given of a solution of Heun's equation expressed in the form of a definite integral or contour integral involving only functions which are, in some sense, simpler} \cite{Ronveaux1}. The Heun equation has 192 different solutions. These solutions were given by Maier only recently \cite{Maier}.

Confluent forms of this equation exist where two regular singular points are coalesced to give us \emph{the confluent Heun equation}. If we coalesce two pairs, we get \emph{the double confluent Heun equation}. If we coalesce three regular singular points, the name given to the differential equation is \emph{the biconfluent Heun equation}. When all four points are coalesced, we obtain \emph{the triconfluent Heun equation}.

We call solutions which are analytic around only one of the singular points \emph{local solutions}. If the solutions are analytic around two singular points, we call them \emph{Heun
functions}. \emph{Polynomial solutions} are analytic around three finite singular points. There are \emph{polynomial solutions} under certain conditions on the parameters of this equation.  Arscott in his article in the book edited by Ronveaux \cite{Arscott} gives conditions to get \emph{polynomial solutions} for Heun type solutions. These occur at discrete values of $q$ and for negative integer $\alpha$. More recent references to these conditions can be found in Fiziev \cite{Fiziev,hakem}, \c{C}ift\c{c}i et al. \cite{Ciftci}, and Karayer et al. \cite{Karayer}.

Reduction of the Heun equation into a simpler equation (i.e. an equation with less singularities) such as the well-known hypergeometric equation is also possible under specific conditions. For a general introduction, reduction to more elementary functions and physical examples, one can see a recent paper\cite{Pelin} and references therein.

Computationally, the computer algebra system Maple has a symbolic and numerical implementation of the Heun's equation and its confluent cases. A recent work of Oleg V. Motygin gives alternative algorithms for the numerical evaluation of the Heun function \cite{Motygin}.
\section{Quantum Field Theory Examples}
Here we want to study two instances where a confluent form of the Heun equation was not recognized. In 1976 't Hooft wrote a seminal paper where he calculated the quantum effects due to a four-dimensional pseudoparticle \cite{Hooft}. In this paper 't Hooft  calculated the one loop contribution to QCD from the BPST instanton \cite{Belavin}. This result was important in those days and was used to get an example of the phase transition to confinement for hadrons \cite{Callan1,Callan2,Callan3,Callan4}.

In his calculation, 't Hooft ended up with an eigenvalue equation, which he could not solve in terms of \emph{simple elementary functions} as he stated on p. 3433 of reference \cite{Hooft}. So, he divided the eigenvalue with a factor $(1+ r^2 )^2$ to get Jacobi polynomial form of the hypergeometric function as a solution with the discrete eigenvalues. He multiplied these discrete eigenvalues to get one loop contribution of the instanton to the QCD. His goal was to calculate the first quantum correction to the value of the tunneling process described by the BPST instanton, using the formula
\begin{equation}
 \mathrm{Det} [\Pi \lambda_i ] =  e^{ \mathrm{Tr} \ln {\Sigma \lambda_i }},
\end{equation}
where $\lambda$ was the eigenvalue of the equation.

He argued that since he would only calculate the ``renormalized" value, if he divided both the term with instanton and the term without the instanton (put there for regularization purposes) he would get the same value for the product of the eigenvalues. Then the new equation reads
\begin{equation}
\frac{d^2u}{dr^2}+ \frac{3}{r}\frac{du}{dr} - \frac{4L^2}{r^2}u -
\frac{4(J_1^2 - L^2)}{1+r^2}u + \frac{4(T^2+\lambda)}{(1+r^2)^2} u = 0.
\end{equation}
This is a hypergeometric equation which will give polynomial solutions with discrete eigenvalues $\lambda_n$. One can consult the original article for the definitions of $L^2$, $J_1^2$ and $T^2$, which we can take as constants in our analysis of the differential equation. For purposes of regularizing, 't Hooft uses Pauli Villars regularization to make the product finite where he also divides the regulator masses by the same factor $(1+r^2)^2$.

The standard form of the confluent Heun equation is given as \cite{Fiziev,hakem}
\begin{equation}
{\frac{{d^{2}w}}{{dz^{2}}}}+\left( \alpha +{\frac{{\gamma+1}}{{z-1}}}+{\frac{{\beta+1}}{{z}}}%
\right) {\frac{{dw}}{{dz}}}+\left( {\frac{{\nu}}{{z-1}}}+{\frac{{\mu}}{{z}}}\right)w =0,
\end{equation}
with solution
\begin{equation}
w=H_C(\alpha, \beta, \gamma, \delta,\eta,z),
\end{equation}
and the parameters have the relations
\begin{equation}
\delta = \mu+\nu-\alpha \bigg( \frac{{ \beta+\gamma+2}}{{2}} \bigg),
\end{equation}
\begin{equation}
\eta = \frac{{ \alpha(\beta+1) }}{{2}} - \mu - \frac {{ \beta+\gamma+ \beta \gamma}}{{2}}.
\end{equation}
One can show that without the extra factors dividing the eigenvalue, the resulting equation can be reduced to the confluent Heun form with a solution given as
\begin{equation}
u = f(r)
H_C \bigg(0,\sqrt{4L^2+1},\sqrt{4T^2+1},-\lambda,-J_1^2+L^2+T^2
+\lambda+\frac{1}{2}, -r^2 \bigg),
\end{equation}
where $H_C$ is the confluent Heun function and $f(r)$ is given by
\begin{equation}
{ f(r) = (-r^2-1)^\frac{\sqrt{4T^2+1}+1}{2}
r^{(-1+\sqrt{4L^2+1})}}.
\end{equation}
We need $f(r)$ to get rid of the quadratic powers of $r$ and $(1+r^2)$ in the denominator of two terms in the differential equation to fit our equation to the form given above.

One wonders if one can find polynomial solutions of this resulting confluent Heun equation. Since in our solution, the first parameter $\alpha$ in Heun vocabulary is zero, we can not find polynomial solutions. This result is stated in the recent \c{C}ift\c{c}i et al. paper \cite {Ciftci}. Karayer et al.\cite{Karayer} give a method that might give a polynomial solution only for zero eigenvalue. According to this method, the condition
\begin{equation}
\mu+\nu=-n \alpha,
\end{equation}
should be satisfied, for non vanishing $\alpha$,  in order to have an $n^{th}$ order polynomial solution. The problem is not being able to set the coefficient of the highest term in the polynomial to zero, even if one can adjust the eigenvalue by equating the proper determinant to zero \cite{Ciftci}. We conclude that we can not find polynomial solutions of the new equation with discrete eigenvalues.

In 1979, seeing this work, one of us (M.H.) tried to do a similar calculation for a Dirac particle coupled to a scalar $CP^1$ model in two dimensions \cite{Hortacsu1}. The $CP^1$ model has an instanton like solution, which can be coupled to the Dirac particle. Mimicking 't Hooft, he divided the eigenvalues by factors of $(1+r^2)$. Then the radial part of the Dirac equation reads
\begin{equation}
\frac{d^2u}{dr^2}+ \frac{1}{r} \bigg( 1 + \frac{2r^2}{1+r^2} \bigg)
\frac{du}{dr} - \bigg[m^2 - \frac{1}{1+r^2} -
\frac{1}{(1+r^2)^2} - \frac{4\lambda^2}{(1+r^2)^2} \bigg] u= 0,
\end{equation}
where $m$ is the angular eigenvalue. As shown in \cite{Hortacsu1}, this equation can be reduced to the hypergeometric equation, with discrete eigenvalues when the series were stopped to result in Jacobi polynomials.

If we do not use the extra factors of $(1+r^2)$ dividing each eigenvalue, the resulting equation reads
\begin{equation}
\frac{d^2u}{dr^2}+ \frac{2m+1}{r} \frac{du}{dr} - \bigg[
\frac{2m+1}{1+r^2} - \frac{3}{(1+r^2)^2} - {4\lambda^2} \bigg] u= 0,
\end{equation}
whose regular solution at $r=0$ is given as
\begin{equation}
u = \frac{ r^m} {\sqrt{-r^2 -1}}  H_C \bigg( 0, m,-2,-\lambda^2, 1+\lambda^2 -\frac{m}{2}, -r^2 \bigg).
\end {equation}
This equation also does not have polynomial solutions since the first parameter in its standard writing is zero.

This is contrary to the case where wave equations are written for particles in the Eguchi-Hanson \cite{Eguchi} and Nutku helicoid instanton \cite{Nutku} metrics. In these two cases whether we can
find a closed system with more elementary functions, or we end up with the Heun class solutions depend on the dimensions of the space we use. As Sucu and \"{U}nal \cite{Sucu} showed, in four dimensions, hypergeometric functions suffice for the solution of wave operators in the background of the Eguchi-Hanson metric. In the same paper, they also showed that a massless Dirac particle in the background of the Nutku helicoid instanton can be solved exactly in terms of exponential functions. One can also expand this solution in terms of an infinite series involving a product of two Mathieu
functions \cite{Cador} if the method of separation of variables is used to obtain the solution. A solution in closed form  as Sucu et al. find is not possible in five dimensions. One can show when one uses the trivially extended metric in five dimensions, the radial equation in the Eguchi-Hanson instanton instanton is solved in terms of confluent Heun functions \cite{Tolga1}. If one trivially extends the Nutku helicoid metric to five dimensions, the wave equation for a massless Dirac particle in this background gives double confluent Heun solutions, which can be reduced to Mathieu functions. Further reduction to more elementary functions is not possible \cite{Tolga,Tolga2}.

Among the people who encounter the Heun type equation and who try to make sense of this equation, without actually solving it, we can cite Cveti\v{c} and Larsen \cite{Cvetic}. They go to the asymptotic region to reduce the Heun equation to the general confluent hypergeometric form (Bessel) in their work. In our second example, if we also go to the asymptotics and drop terms of higher order in $1/r$, we get the Bessel equation. This is no surprise, since by going to the asymptotics, we have ignored all the interaction terms, reducing our equation to the free wave case. These equations, also, do not have discrete eigenvalues unless we put the system into a box.
\section{Conclusion}
Here we introduced Heun equation and its confluent forms and gave two examples where the confluent Heun equation was not properly identified in quantum field theory literature in 1970s.  We show that 't Hooft's eigenvalue equation emerging in the calculation of the one loop contribution to QCD from the BPST instanton and Horta\c{c}su's calculation for a Dirac particle coupled to a scalar $CP^1$ model in two dimensions can be given in terms of confluent Heun equations in their unaltered forms. We then show that the equations in their original forms can not have polynomial solutions with discrete eigenvalues.

We should also emphasize that 't Hooft used his remarkable insight to give a meaning to an equation, which is not solvable in a closed form by normal means. Since the original equation does not have discrete eigenvalues, it would not served 't Hooft's purpose even if he could solve it exactly. He was only trying to calculate the first quantum correction to a physical process, and was not interested in the properties of the differential equation. He needed discrete eigenvalues for this purpose.  At the end he achieved his goal to calculate the ``the first quantum correction to the value of the tunneling process described by the BPST instanton", the physically important result, by modifying the equation, without changing the physical result. If 't Hooft did not use his remarkable foresight to reduce his equation in his way to get the final regularized result, he might not been able to calculate his celebrated result.
\section*{Acknowledgement}
M.H. thanks Dr. Nadir Ghazanfari for technical assistance. He also thanks Science Academy, Turkey, for support. This research is also supported by TUBITAK, the Scientific and Technological Council of Turkey.


\begin{thebibliography}{99}
\bibitem{Heun} K.~Heun: \emph{Math. Ann.} \textbf{33}, 161 (1889).

\bibitem {Ronveaux3} A.~Ronveaux (ed.): \emph{Heun's Differential Equation}, Oxford University Press, Oxford 1995.

\bibitem{Slavyanov} S.~Y.~Slavyanov and S.~Lay: \emph{Special Functions, A Unified Theory Based on Singularities}, Oxford University Press, Oxford 2000.

\bibitem {Hortacsu} M.~Horta\c{c}su: \emph{Heun Functions and their uses in Physics} in Proceedings of the 13th Regional Conference on Mathematical Physics, Antalya, Turkey, October 27-31, 2010, U. Camc{\i} and I. Semiz eds., p. 23, World Scientific, Singapore 2013.

\bibitem {Teukolsky} S.~A.~Teukolsky: \emph{Phys. Rev. Lett.} \textbf{29}, 1114 (1972).

\bibitem {Batic} D.~Batic and H.~Schmid: \emph{J. Math. Phys.} \textbf{48}, 042502 (2007).

\bibitem {Poons} J.~Blaudin, R.~Poons and G.~Marcilhacy: \emph{Lett. Nuovo Cim.} \textbf{38}, 561 (1983).

\bibitem {Leaver} E.~W.~Leaver: \emph{J. Math. Phys.} \textbf{27}, 1238 (1986).

\bibitem {Suzuki} H.~Suzuki, E.~Takasugi and H.~Umetsu: \emph{Prog. Theor. Phys.} \textbf{100}, 491 (1998).

\bibitem {Nutku} A.~N.~Aliev, M.~Horta\c{c}su, J.~Kalayc\i\ and Y.~Nutku: \emph{Class. Quant. Grav.} \textbf{16}, 631 (1999).

\bibitem {Tolga} T.~Birkandan and M.~Horta\c{c}su: \emph{J. Math. Phys.} \textbf{48}, 092301 (2007).

\bibitem {Tolga2} T.~Birkandan and M.~Horta\c{c}su: \emph{J. Phys. A} \textbf{40}, 1105 (2007).

\bibitem {Hooft} G.~'t Hooft: \emph{Phys. Rev. D} \textbf{14}, 3432 (1976).

\bibitem {Belavin} A.~A.~Belavin, A.~M.~Polyakov, A.~S.~Schwartz and Yu.~S.~Tyupkin: \emph{Phys. Lett. B} \textbf{59}, 85 (1975).

\bibitem {Hortacsu1} M.~Horta\c{c}su: \emph{Phys. Rev. D} \textbf{20}, 496 (1979).

\bibitem{Arscott} F.~M.~Arscott: \emph{Heun's Equation} in Heun's Differential Equation, p. 41, A. Ronveaux ed., Oxford University Press, Oxford 1995.

\bibitem{Fiziev} P.~P.~Fiziev: \emph{Class. Quant. Grav.} \textbf{27}, 135001 (2010).

\bibitem{hakem} P.~P.~Fiziev: \emph{J. Phys. A: Math. Theor.} \textbf{43}, 035203 (2010).

\bibitem{Ciftci} H.~\c{C}ift\c{c}i, R.~L.~Hall, N.~Saad and E.~Do\u{g}u: \emph{J. Phys. A: Math. Theor.} \textbf{43}, 415206 (2010).

\bibitem {Karayer} H.~Karayer, D.~Demirhan and F.~B\" {u}y\" {u}kk{\i}l{\i}\c{c}: \emph{J. Math. Phys.} \textbf{56}, 063504 (2015).

\bibitem{Ronveaux1} F.~M.~Arscott: \emph{Heun's Equation} in Heun's Differential Equation, p. 65, A. Ronveaux ed., Oxford University Press, Oxford 1995.

\bibitem{Maier} R.~S.~Maier: \emph{J. Differential Equations} \textbf{213} 171 (2005).

\bibitem{Pelin} P.~Aydiner and T.~Birkandan: \emph{Physical problems admitting Heun-to-hypergeometric reduction} in Proceedings of the International Conference DAYS on DIFFRACTION 2015, p. 27, O.V. Motygin, A.P. Kiselev, P.A. Belov, L.I. Goray, A.Ya. Kazakov and A.S. Kirpichnikova eds., IEEE, St. Petersburg 2015.

\bibitem{Motygin} O.~V.~Motygin: \emph{On evaluation of the Heun functions} in Proceedings of the International Conference DAYS on DIFFRACTION 2015, pp. 222-228, O.V. Motygin, A.P. Kiselev, P.A. Belov, L.I. Goray, A.Ya. Kazakov and A.S. Kirpichnikova eds., IEEE, St. Petersburg 2015.

\bibitem {Callan1} C.~G.~Callan, R.~H.~Dashen and D.J.Gross: \emph{Phys. Rev. D} \textbf{17}, 2717 (1978).

\bibitem {Callan2} C.~G.~Callan, R.~H.~Dashen and D.~J.~Gross: \emph{Phys. Rev. D} \textbf{19}, 1826 (1979).

\bibitem{Callan3} C.~G.~Callan, R.~H.~Dashen and D.~J.~Gross: \emph{Phys. Rev. D} \textbf{20}, 3279 (1979).

\bibitem {Callan4} C.~G.~Callan, R.~H.~Dashen, D.~J.~Gross, F.~Wilczek and A.~Zee: \emph{Phys. Rev. D} \textbf{18}, 4684 (1978).

\bibitem {Eguchi} T.~Eguchi and A.~J.~Hanson: \emph{Phys. Lett. B} \textbf{74}, 249 (1978).

\bibitem {Sucu} Y.~Sucu and N.~\" {U}nal: \emph{Class. Quant. Grav.} \textbf{21}, 1443 (2004).

\bibitem {Cador} L.~Chaos-Cador and E.~Ley-Koo: \emph{Rev. Mexicana F{\'i}s.} \textbf{48}, 67 (2002).

\bibitem{Tolga1} T.~Birkandan and M.~Horta\c{c}su: \emph{J. Math. Phys.} \textbf{49}, 054101 (2008).

\bibitem{Cvetic} M.~Cveti\v{c} and F.~Larsen: \emph{Phys. Rev. D} \textbf{56}, 4994 (1997).

\end{thebibliography}
\end{document}